# Abrupt Periodic Pulsation Resumptions in Deneb

*Joyce A. Guzik[1], Helmut A. Abt[2], Jason Jackiewicz[3], and Brian Kloppenborg[4]*

[1]Los Alamos National Laboratory, Los Alamos, NM 87545; joy@lanl.gov
[2]Kitt Peak National Observatory, Box 26732, Tucson, AZ 85712
[3]New Mexico State U., Las Cruces, NM 88003
[4]AAVSO, 185 Alewife Brook Parkway, Suite 410, Cambridge, MA 02138

**Subject Keywords**

AAVSO International Database; Photometry, CCD; Radial Velocities, Alpha Cygni Variables; stars: individual (alpha Cyg)

**Abstract**

Deneb ($\alpha$ Cygni) is a bright (V magnitude 1.25) blue-white supergiant (spectral type A2 Ia) which shows variability in both radial velocity and photometric measurements. H. Abt reviewed radial velocity measurements by Paddock (1935) using the Lick observatory 36-inch telescope spectrograph during 1927-1935. Abt noticed resumptions of pulsations with a dominant quasi-period of around 12 days that occur at intervals of around 70 days, and damp out after a few cycles. These resumptions appear to happen at arbitrary phase. Perhaps another event like this was captured in a shorter series of radial velocity measurements by Abt in 1956. We examined subsequent radial velocity and photometric data available in the literature, along with photometric measurements by the TESS spacecraft and V-magnitude observations by AAVSO observers. We find some evidence for periodic resumptions of larger-amplitude pulsations in these data. However, longer contiguous data sets combined with more frequent sampling are needed to confirm this periodicity in resumption of pulsations, and to help answer many more questions about Deneb and the $\alpha$ Cygni variables.

**1. Introduction**

Despite being one of the brightest stars of the Northern Hemisphere sky (apparent visual magnitude 1.25), surprisingly little is known about Deneb (Diodati 2019). Deneb is the prototype of the $\alpha$ Cyg variables (Abt 1957, van Genderen et al. 1989), which show small-amplitude (~0.1 mag) irregular variations. Deneb varies in both radial velocity and brightness with dominant period around 12 days (Paddock 1935). Its effective temperature is 8500 K and spectral type A2 Ia (Schiller and Przybilla 2008). Deneb's current mass is estimated at about 20 solar masses. Deneb's mass-loss rate is 3 x 10$^{-7}$ solar masses/year (Schiller and Przybilla 2008), 10 million times higher than that of the Sun, amounting to two Mercury masses per year.

The distance to Deneb is controversial (see, e.g., Diodati 2019, Byrd and Machholz 2023). Hipparcos parallaxes have large uncertainties, and instead the distance to Deneb of 2600 light years is based on the distance to stars in the Cygnus OB7 association, assuming that Deneb is a member. Deneb's luminosity using this distance is around 200,000 solar luminosities (Schiller and Przybilla 2008). Note that a more definitive Gaia parallax for Deneb is unavailable (as yet) because Deneb is so bright that it would saturate Gaia's detectors (Byrd and Machholz 2023). Deneb's radius, based on its luminosity and temperature, is 200 solar radii. If Deneb were at the location of the Sun, it's surface would nearly reach Earth's orbit, since the Earth-Sun distance is 212 solar radii.

Deneb's age and evolutionary state are also not known—it may be on its first crossing of H-R diagram on its way to becoming a red supergiant before reaching core helium-burning phase, or it may be on its way back toward the blue. According to Abt et al. (2023), there is evidence that $\alpha$ Cyg stars are on their first crossing of the HR diagram, because at least one such star (6 Cas) is a member of the Cas OB5 association (Bartaya et al. 1994). Gorlova et al. (2006) derived a young age of 4-6 Myr for that association. However, Saio et al. (2013) find, using stellar evolution and pulsation modeling, that $\alpha$ Cyg-like pulsations are excited only after the star has been a red supergiant. On the other hand, they note that the N/C and N/O abundance ratios for Deneb are more consistent with stars that have not yet reached the red supergiant stage.

Deneb's projected rotational velocity (v sin i) is only around 20 km/sec, but interferometric measurements show 2% asymmetry in the disk, indicating that we could be observing Deneb close to pole-on (inclination 30° with respect to our line of sight), and that the rotation velocity could be as high as 35% of breakup (Aufdenberg et al. 2008). However, there are ambiguities in the data, and more work is required to confirm the fast-rotator interpretation (Chesneau et al. 2010).

Deneb is located among the Luminous Blue Variables in the H-R diagram (Humphreys and Davidson 1994), but doesn't show LBV outbursts, i.e., abrupt large mass ejections and excursions to the red in the H-R diagram that occur decades apart (see, e.g., Guzik and Lovekin 2012). Deneb's low amplitude variability, however, may have the same origin as LBV microvariations.

**2. Radial Velocity Measurements**

Abt et al. (2023) re-examined an extensive series of radial velocity measurements made by Paddock (1935; see also Lucy 1976) using the Mills three-prism spectrograph on the 36-inch Lick Observatory refractor from 1927 through 1935. Abt et al. notice that the radial velocities show abrupt resumptions of larger-amplitude pulsations which are uncorrelated with pulsation phase. They mark several instances where these events occur at points A, B, C, and D (Fig. 1).

Abt et al. (2023) also notice that the time interval between points A to B and between points B to C is around 70 days, and that the interval between C and D is a multiple of about 70 days (see Table 1). Another abrupt amplitude increase may have been observed by Abt in 1956 (Abt 1957), labeled as point E in Figure 2. The data around point E were taken 25 years later than the Paddock data, but are not inconsistent with a multiple of around 70 days.

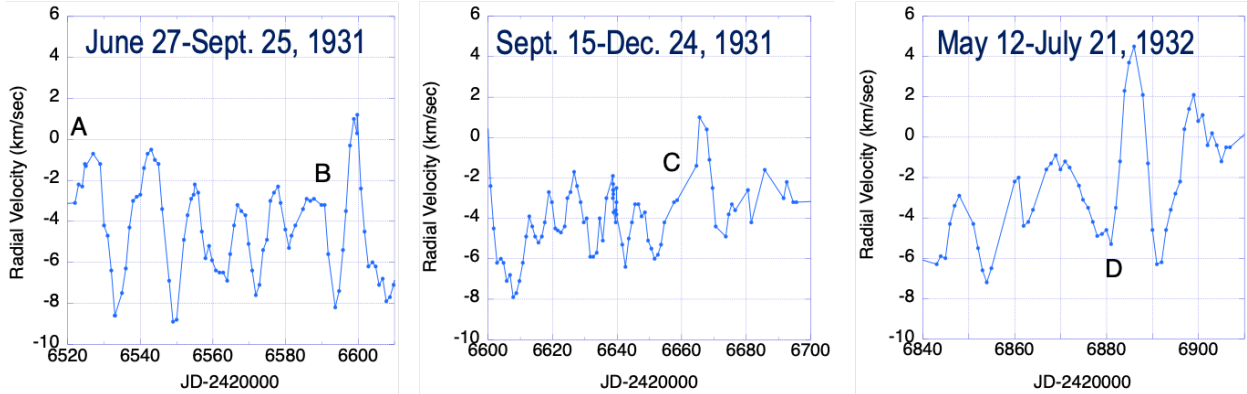

*Figure 1. Deneb Radial velocity vs. Julian date from Paddock et al. (1935). Instances of abrupt resumption of larger amplitude pulsations are marked by letters A, B, C, and D.*

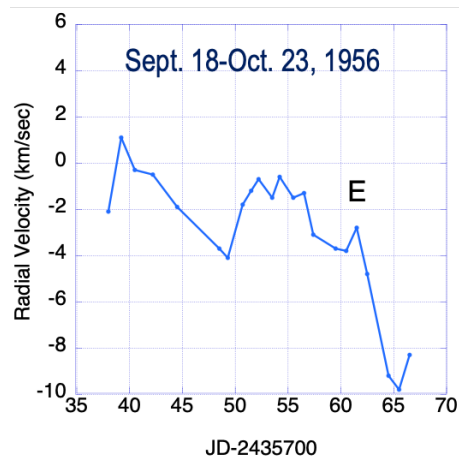

*Figure 2. Deneb radial velocity vs. Julian Date from observations by Abt (1957). Another such resumption of larger amplitude pulsation may have occurred near the date labeled at E.*

Table 1. Time intervals between radial velocity points A through E in Figures 1 and 2.

| Point | Julian Date | Interval (days) | Possible Multiple |
|---|---|---|---|
| A | 2426517 ± 5 | | |
| | | 75 ± 5 | 75 days x 1 |
| B | 2426592 ± 0.5 | | |
| | | 69 ± 4 | 69 days x 1 |
| C | 2426661 ± 3 | | |
| | | 220 ± 4 | 73 days x 3 |
| D | 2426881 ± 0.5 | | |
| | | 8866 ± 1 | 71 days x 125 |
| E | 2435747 ± 1 | | |

### 3. Later Deneb Time-Series Data

The referee of Abt et al. (2023) requested that we examine more recent archival data in photometry and radial velocity to search for additional instances of abrupt pulsation resumptions. The referee in particular suggested Transiting Exoplanet Survey Satellite (TESS, Ricker et al. 2015) data and Hipparcos spacecraft (van Leeuwen 1997) data. Our search also turned up additional ground-based observations.

Parthasarathy and Lambert (1987) observed Deneb using the 2.7-meter reflector and coudé spectrometer at McDonald Observatory between March 1, 1980 and November 26, 1982. Because Deneb is so bright they could conduct either daytime or night-time observations. The radial velocity measurements derived from these observations (Figure 3, 213 points) show a few possible large excursions. The A-B interval is 66 days. The B-C interval is 412 days = 6 x 69 days.

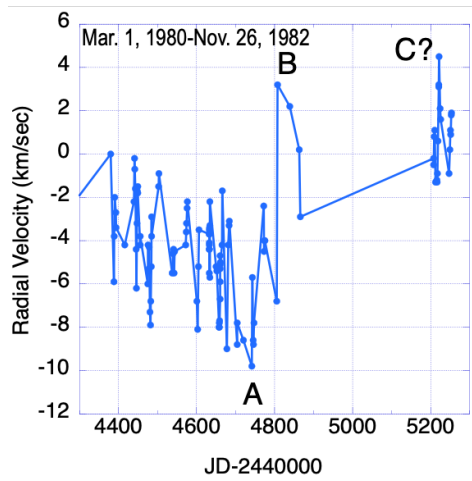

*Figure 3. Deneb radial velocity vs. Julian Date from Parthasarathy and Lambert (1987).*

The Hipparcos photometry is too sparse to identify abrupt brightness excursions. Data points exist for only 30 distinct days over four years, March 3, 1990 through February 23, 1993. However, Hipparcos spacecraft data were used to discover 32 $\alpha$ Cyg variables (Waelkens et al. 1998).

Richardson et al. (2011) present Deneb Strömgren v-magnitude photometry and radial-velocity measurements from spectroscoposcopic observations taken April 1997 through December 2001. The spectroscopy was done using a 1-meter telescope and echelle spectrograph at Ritter Observatory, and the photometry taken using the 0.75-meter Four College Automated Photoelectric Telescope (FCAPT). There are 368 photometry data points, and 281 radial-velocity data points. We plotted their data in Figure 4 and labeled the largest excursions from the mean magnitude and radial velocity by points A, B, C, and D. Point C seems to be near an excursion in both photometry and radial velocity. We find three nearly equal intervals between these events of 447-448 days. These intervals would not be inconsistent with six 75-day intervals or seven 64-day intervals. The data are too sparse to reliably identify or rule out pulsation resumption events between points A, B, C, and D.

Figure 5 zooms in on the Richardson et al. (2011) photometric data in Fall of 1998 and Fall of 1999 which have the most frequent coverage. The characteristic 12-day quasi-periodicity is evident in these light curves.

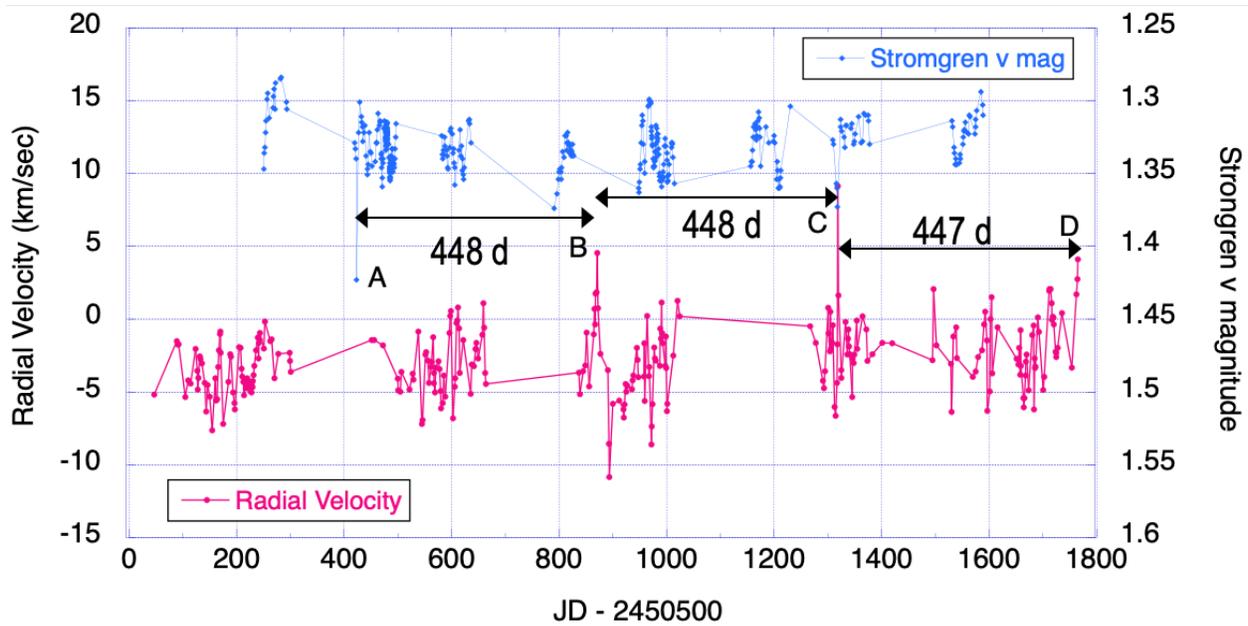

*Figure 4. Deneb radial-velocity and photometry data from Richardson et al. (2011). Four large excursions were identified at A, B, C, and D, which are nearly equally spaced at 447-448 days.*

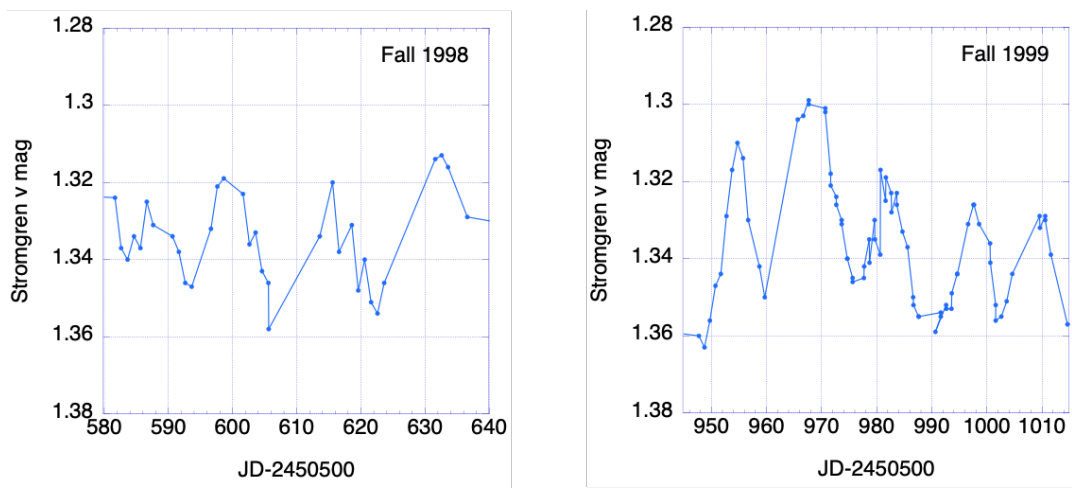

*Figure 5. Zoom-in on Richardson et al. (2011) Deneb photometry from Fall 1998 and Fall 1999, showing variability with periodicity around 12 days.*

The Mikulski Archive for Space Telescopes (MAST, https://archive.stsci.edu) contains High-Level Science Product (HLSP) light curves for Deneb in 2-minute cadence photometry taken during three 27-day sectors. After removing a few outlying points, we plotted the normalized Simple Aperture Photometry (SAP) flux measurements vs. time in Figure 6. Note that we did not use the Pre-Data Conditioned SAP fluxes, in which long-period trends, including some pulsations, have been smoothed out. These observations show irregular pulsations during Sector 41 and resumption of higher amplitude pulsations with an approximate 12-day period during Sector 55, continuing into Sector 56.

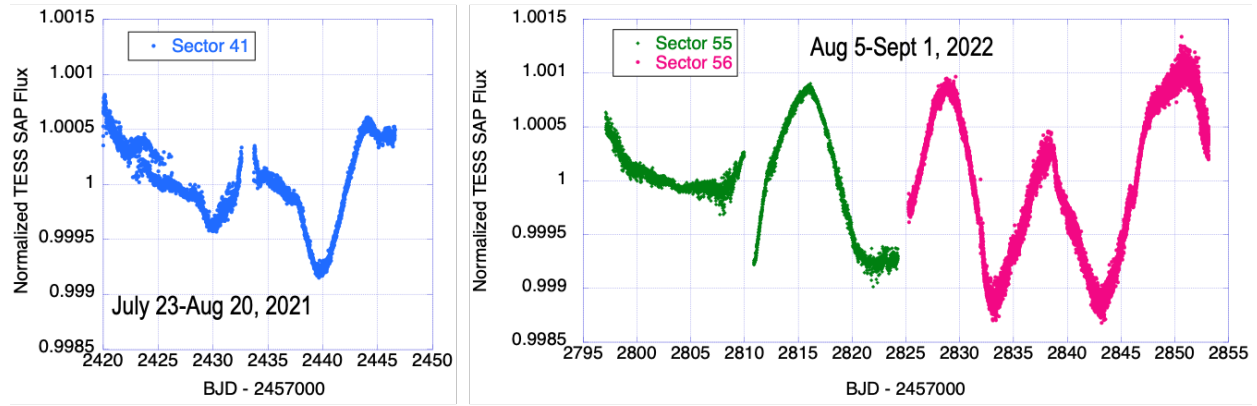

*Figure 6. Deneb light curves from TESS data during Sectors 41, 55, and 56. Large-amplitude pulsations appear to resume during Sector 55 at around day 2810 after BJD 2457000.*

## 4. Recent Deneb AAVSO Data

We also examined the American Association of Variable Stars International Database (AID, Kloppenborg 2023) for data on Deneb, and were pleased to find a recent extensive set of V-band photometry taken by ten different observers during the past two years, June 16, 2021–June 15, 2023, comprising 128 data points. The AAVSO observer codes are CTOA, FBA, FXJ, MPFA, BVE, DFR, BWU, GMV, GTIA, and WIG. Figure 7 shows V magnitude vs. time and Figure 8 zooms in on two time spans with the densest coverage, in which Deneb's 12-day quasi-periodicity can be discerned. There appears to be at least one large excursion from mean magnitude around day 200 after JD-2459300, but the time-series coverage is a little too sparse to reliably identify the date of additional abrupt amplitude increases.

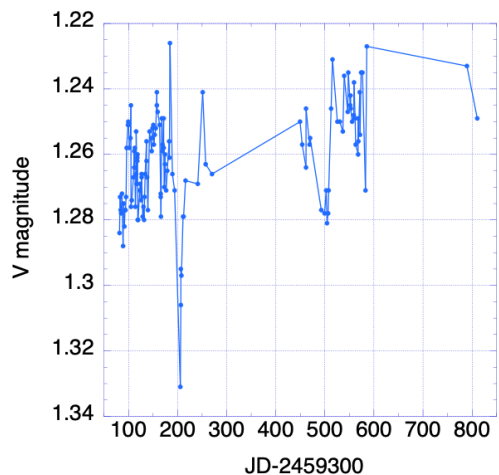

*Figure 7. Deneb V magnitude vs. Julian Date from the AAVSO International Database (AID). These data were taken between June 16, 2021 and June 15, 2023.*

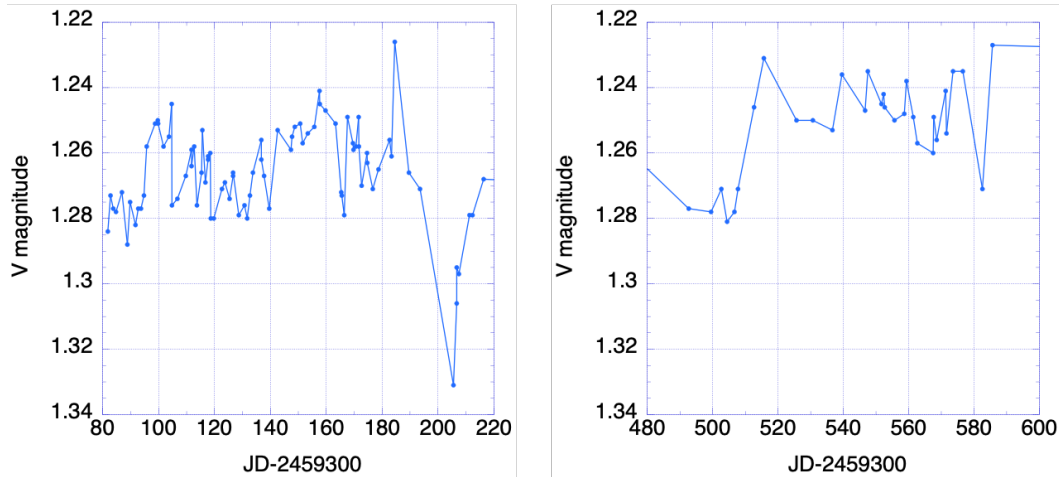

*Figure 8. Zoom-in on portions of Deneb AID data showing some large excursions in brightness and approximate 12-day pulsation periods.*

## 5. Conclusions

We have many questions about Deneb's variability that additional observations may help to answer: Can we confirm that pulsation resumptions occur at around 70-day intervals, and how precise are these intervals? Do these resumptions occur at an arbitrary phase in the 12-day dominant quasi-period? Do they occur at the same time in light curve and radial velocity data? Are there additional radial or non-radial pulsation modes with well-defined periods and amplitudes? Note that Lucy (1976) reports finding many periodicities in a Fourier analysis of the Paddock (1935) data, attributed to nonradial pulsations, but our analysis did not find any significant periods.

The available TESS time series is too short/discontinuous and the AAVSO ground-based photometry data points are a little too infrequent to answer these questions definitively. However, several additional 27-day sectors of TESS data, preferably contiguous, as well as AAVSO time series data with around one data point per night for several consecutive 70-day intervals may be possible to obtain. Perhaps multi-site observations of Deneb using the Bright Star Monitor network would be useful. Perhaps it is also possible for observers to measure radial velocities to within 1-2 km/sec with cadence around one data point per night using spectroscopy as did Paddock the early 1930s. It would also be worthwhile to investigate other $\alpha$ Cygni variables to see whether they show any similarities in behavior.

Deneb inspires many questions and challenges for stellar modeling:

- What is the evolutionary state of Deneb?
- What is the cause of its pulsations with quasi-period 12 days? Why 12 days?
- Why do the large-amplitude pulsations damp out and resume, perhaps every 70 days?
- Are there multiple pulsation modes occurring simultaneously?
- Does Deneb have properties in common with other $\alpha$ Cygni variables?

- What distinguishes Deneb from the Luminous Blue Variables in the same part of the H-R diagram?

**6. Epilogue**

During the question-and-answer session following our November 4 AAVSO Annual Meeting talk, AAVSO Director Brian Kloppenborg offered unpublished photometric data on Deneb from the Solar Mass Ejection Imager taken from 2003 through 2011 (see, e.g., Jackson et al. 2004, Clover et al. 2011). The spacecraft was in a Sun-synchronous polar orbit with a 102-minute period. While the spacecraft was designed to measure solar coronal mass ejections (CMEs), it also observed almost every bright (V < 6) star. It had three cameras equipped with non-linear optics that provided each camera with a 3 x 120 degree field-of-view of the sky.

After the meeting, Brian located these data for Deneb (Figures 9 and 10). These data require further processing to remove camera-to-camera zero-point offsets, angle-dependent flux loss, and other spacecraft-related noise and artifacts. The angle-dependent flux loss causes the ~100-day curvature or changes in slope. The light-curve cadence is around 102 minutes (once per spacecraft orbit), and the time series is long enough that it may be definitive to answer some of our questions about Deneb. Developing a processing pipeline for the SMEI data will be valuable for other variable-star projects since SMEI collected light curves for around 6500 bright stars.

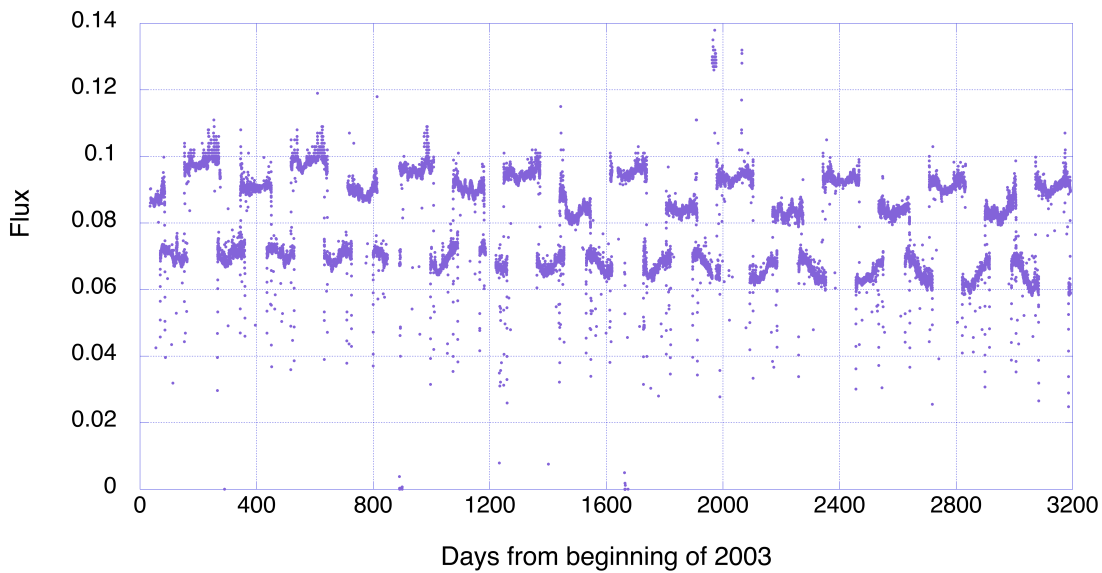

*Figure 9. Deneb time-series uncorrected data from Solar Mass Ejection Imager (Kloppenborg, private communication).*

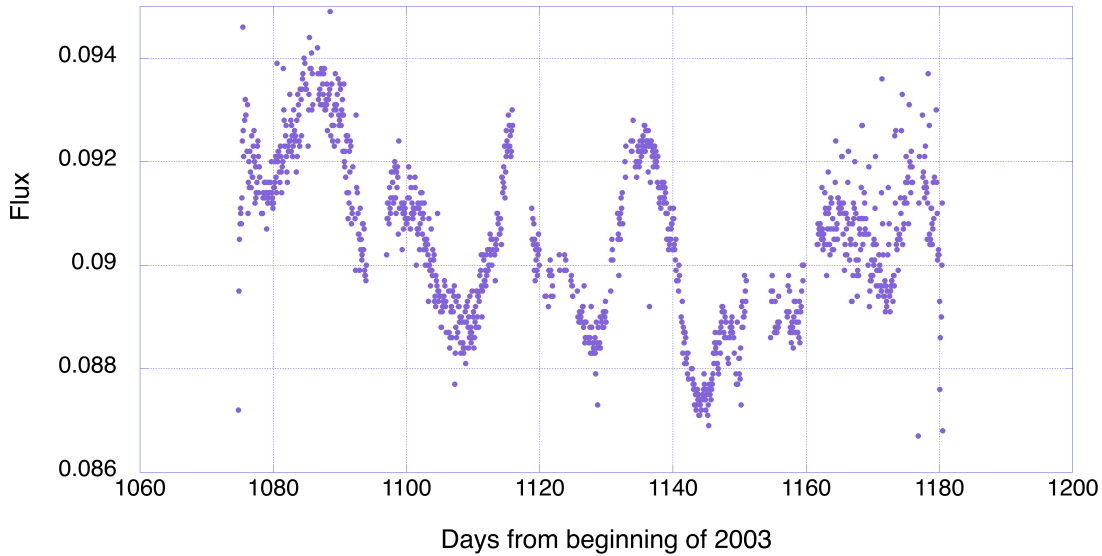

*Figure 10. Zoom-in on portion of uncorrected Deneb data from Solar Mass Ejection Imager.*

We found that there are 101 α Cygni variables listed in the AAVSO Variable Star Index (VSX, https://www.aavso.org/vsx/, Watson et al. 2006) database. These include several bright well-known stars, for example: Rigel and Saiph at the lower left and right corners of Orion; Alnilam, the star in the center of Orion's belt; and Aludra, the star at the tail of Canis Major. This database will be a useful starting point for comparing Deneb to other α Cyg variables and motivating additional data searches and observations.

**Acknowledgements**

This research made use of data from the Mikulski Archive for Space Telescopes (MAST). We acknowledge with thanks the variable star observations from the AAVSO International Database contributed by observers worldwide and used in this research. We thank Jos de Bruijne from the ESA Hipparcos support center for obtaining α Cyg photometry from the Hipparcos archive. This collaboration was facilitated by a Los Alamos National Laboratory Center for Space and Earth Sciences grant XX8P ASF2. J.G. acknowledges support from Los Alamos National Laboratory, managed by Triad National Security, LLC for the U.S. DOE's NNSA, Contract #89233218CNA000001. We thank John Dragon for carefully proofreading this manuscript.